\documentclass[%
reprint,
superscriptaddress,
 amsmath,
 amssymb,
 aps,
 amsfonts,
prl
]{revtex4-2}

\usepackage{graphicx}
\usepackage{dcolumn}
\usepackage{bm}
\usepackage{physics}
\usepackage{hyperref}
\usepackage{soul}

\hypersetup{
    colorlinks=true,
    linkcolor=blue,
    filecolor=magenta,      
    urlcolor=blue,
    citecolor=blue,
    pdftitle={Overleaf Example},
    pdfpagemode=FullScreen,
    }

\begin{document}
\title{Gauge theory of giant phonon magnetic moment in doped Dirac semimetals}
\author{Wenqin Chen}
\affiliation{Department of Physics, University of Washington, Seattle, Washington 98195, USA}
\affiliation{Theoretical Division, T-4 and CNLS, Los Alamos National Laboratory, Los Alamos, New Mexico 87545, USA}
\author{Xiao-Wei Zhang}
\affiliation{Department of Materials Science and Engineering, University of Washington, Seattle, Washington 98195, USA}
\author{Ying Su}
\affiliation{Center for Integrated Nanotechnologies (CINT), Los Alamos National Laboratory, Los Alamos, New Mexico 87545, USA}
\author{Ting Cao}
\affiliation{Department of Materials Science and Engineering, University of Washington, Seattle, Washington 98195, USA}
\author{Di Xiao}
\email{dixiao@uw.edu}
\affiliation{Department of Materials Science and Engineering, University of Washington, Seattle, Washington 98195, USA}
\affiliation{Department of Physics, University of Washington, Seattle, Washington 98195, USA}
\author{Shi-Zeng Lin}
\email{szl@lanl.gov}
\affiliation{Theoretical Division, T-4 and CNLS, Los Alamos National Laboratory, Los Alamos, New Mexico 87545, USA}
\affiliation{Center for Integrated Nanotechnologies (CINT), Los Alamos National Laboratory, Los Alamos, New Mexico 87545, USA}

\begin{abstract}
    We present a quantum theory of phonon magnetic moment in doped Dirac semimetals. Our theory is based on an emergent gauge field approach to the electron-phonon coupling, applicable to both gapless and gapped systems. We find that the magnetic moment is directly proportional to the electrical Hall conductivity through the phonon Hall viscosity. Our theory is combined with the first-principles calculations, allowing us to quantitatively implement it to realistic materials. Magnetic moments are found to be on the order of Bohr magneton for certain phonon modes in graphene and $\text{Cd}_3 \text{As}_2$. Our results provide practical guidance for the dynamical generation of large magnetization in the topological quantum materials.
\end{abstract}

\maketitle

\textit{Introduction}.--- Circularly polarized phonons, the collective excitations of ionic circular motions \cite{Chiral-Phonons,Observation-of-chiral,Truly-chiral,Chiral-phonons-in-quartz}, have recently attracted significant interest due to their contributions to various phenomena such as the Einstein-de Haas effect \cite{Angular-Momentum,The-ultrafast-Einstein}, the thermal Hall effect \cite{Giant-thermal,Phonon-Thermal,Chiral-phonons-in-the,Phonon-Angular-Momentum-Hall,Thermal-Hall-conductivity,Large-phonon-thermal-Hall,Berry-Phase-of-Phonons,Phonon-Hall-Viscosity}, the spin-phonon angular momentum transfer \cite{Frequency-Splitting-of-Chiral-Phonons,Adiabatic-Dynamics-of-Coupled-Spins,Conversion-between-electron-spin} and phonon-induced effective magnetic fields \cite{An-effective-magnetic-field,Large-effective-magnetic,Terahertz-electric-field,Phononic-switching-of-magnetization,Phono-magnetic-analogs,Giant-effective-magnetic-fields,Phonon-driven-spin-Floquet,Light-Driven-Spontaneous,Dynamically-induced-magnetism}. These phonons carry an orbital magnetic moment, classically understood as a ionic loop current of Born effective charge \cite{Dynamical-multiferroicity,Orbital-magnetic}. The magnitude is predicted to be on the order of the nuclear magneton $\mu_N$. Phonon magnetic moments have been observed in experiments via the phonon Zeeman effect across several materials \cite{A-Large-Effective,Magnetic-Control,Observation-of-interplay}. Surprisingly, the measured moments can be up to a few Bohr magneton $\mu_B$, orders of magnitude larger than the classical prediction, indicating the necessity of quantum theories capturing the contributions from electronic degrees of freedom.

Towards this goal, a quantum theory has been developed from the adiabatic pumping of electronic current for band insulators \cite{Geometrodynamics-of-electrons,Geometric-orbital-magnetization,Phonon-Magnetic-Moment,Gate-Tunable-Phonon}. However, it diverges when the band gap closes due to the breakdown of the adiabatic approximation, and therefore is unable to handle the metallic phase of materials \cite{Phonon-Magnetic-Moment}. More recently, a microscopic theory based on the orbit-lattice coupling has been proposed for magnetic materials \cite{Giant-effective-magnetic-moments}. On the other hand, a phenomenological model of the phonons coupled to the cyclotron motion of carriers has been used for Dirac semimetals \cite{A-Large-Effective}, however, the microscopic mechanism is still unclear. In fact, a theory that quantitatively accounts for the gapless systems remains absent.

In this Letter, we propose a quantum theory for the phonon magnetic moment in doped Dirac semimetals. Our theory is based on an emergent gauge theory approach to the electron-phonon coupling.  The mechanism of time-reversal symmetry (TRS) breaking is formulated in a topological Chern-Simons term, which appears as a phonon Hall viscosity modifying the phonon dynamics. We find that the phonon magnetic moment is directly linked to the Hall conductivity through the phonon Hall viscosity. Our results provide a theoretical framework to calculate the phonon magnetic moment from the basic properties of crystal structure and electronic transport.  We then apply our theory to realistic materials such as graphene and $\text{Cd}_3 \text{As}_2$ by establishing a first-principles method for the computation of phonon-induced emergent gauge fields. Giant phonon magnetic moments on the order of $\mu_B$ are found for certain phonon modes. Our theory serves as practical guidance for the dynamical generation of large magnetization in materials. 

\textit{Electron-phonon coupling and emergent gauge fields in Dirac semimetals}.--- We start from a general model for Dirac semimetals with electron-phonon ($e$-ph) coupling, described by the Hamiltonian $\mathcal{H} = \mathcal{H}_D + \mathcal{H}_{e\text{-ph}}$. Here we consider Dirac semimetals with two valleys ($\boldsymbol{K}_\chi$, $\chi = \pm$) located away from the $\Gamma$ point, which are related by time-reversal or inversion. The low-energy Hamiltonian at one valley is written as $\mathcal{H}_D =\hbar \sum_j v_j k_j \gamma^j - \varepsilon_F$ where  $v_j$ is the Fermi velocity, $\gamma^j$ the Dirac matrices, and $\varepsilon_F$ the Fermi energy~\cite{RevModPhys.90.015001}. The $e$-ph coupling Hamiltonian generally has the form \cite{Mahan,RevModPhys.89.015003}
\begin{equation}
    \mathcal{H}_{e\text{-ph}}^{\nu} = \sum_{\boldsymbol{q}}\sum_{\alpha\beta} g^{\nu}_{\alpha\beta}(\boldsymbol{k},\boldsymbol{q})Q^{\nu}_{\boldsymbol{q}} c^\dagger_{\alpha,\boldsymbol{k}+\frac{\boldsymbol{q}}{2}} c_{\beta,\boldsymbol{k}-\frac{\boldsymbol{q}}{2}},
\end{equation}
where $g^{\nu}_{\alpha\beta}(\boldsymbol{k},\boldsymbol{q})$ is the $e$-ph coupling matrix element,  $\nu$ labels phonon modes, $\boldsymbol{q}$ is the phonon wavevector, and $\alpha,\beta$ are indices for the electronic basis. The phonon displacement operator is defined in terms of the bosonic operators as $Q^{\nu}_{\boldsymbol{q}} = [\hbar/(2 m_I \omega_{\boldsymbol{q}}^{\nu})]^{1/2} (b^{\nu}_{\boldsymbol{q}}+b^{\nu\dagger}_{-\boldsymbol{q}})$ where $m_I$ is the ionic mass and $\omega_{\boldsymbol{q}}^{\nu}$ is the mode frequency. Next we project $\mathcal{H}^{\nu}_{e\text{-ph}}$ into the basis of massless Dirac fermions at the $\boldsymbol{K}_\chi$ valley. The lowest-order coupling, for which $g^{\nu}_{\alpha\beta}(\boldsymbol{q})$ is independent of $\boldsymbol{k}$, can lead to the emergence of $U(1)$ gauge fields if $g^{\nu}_{\alpha\beta}(\boldsymbol{q})$ is compatible with the little group symmetries at the Dirac point $\boldsymbol{K}_\chi$ \cite{Phonons-and-electron-phonon,Symmetry-based,VOZMEDIANO2010109,Elastic_Gauge,Chiral-Anomaly,Inhomogeneous-Weyl-and-Dirac-Semimetals,Phonon_Helicity}. The emergent gauge field interacts with Dirac fermions in the form of minimal coupling as described by an effective Hamiltonian,
\begin{equation}
\label{hamiltonian}
    \mathcal{H}_{\text{eff}} = \sum_j v_j (\hbar k_j -e A_j - e \chi a_j^{\nu}) \gamma^j - \varepsilon_F,
\end{equation}
where $\boldsymbol{A}$ is the electromagnetic(EM) gauge field and $\boldsymbol{a}^{\nu}$ is the emergent gauge field induced by the phonon mode $\nu$. We note that the emergent gauge fields at the two valleys ($\chi=\pm$) have opposite signs as required by the TRS.  In a more general context, the emergent gauge field must transform equivalently as $\boldsymbol{k}$ under the little group at $\boldsymbol{K}_\chi$ for the minimal coupling to be allowed. It is worth mentioning that there can be an additional scalar field induced by the phonons. It affect $\varepsilon_F$ as an electrostatic pseudopotential and thus will not be our primary focus here.

\begin{figure}
    \centering
    \includegraphics[width=\linewidth]{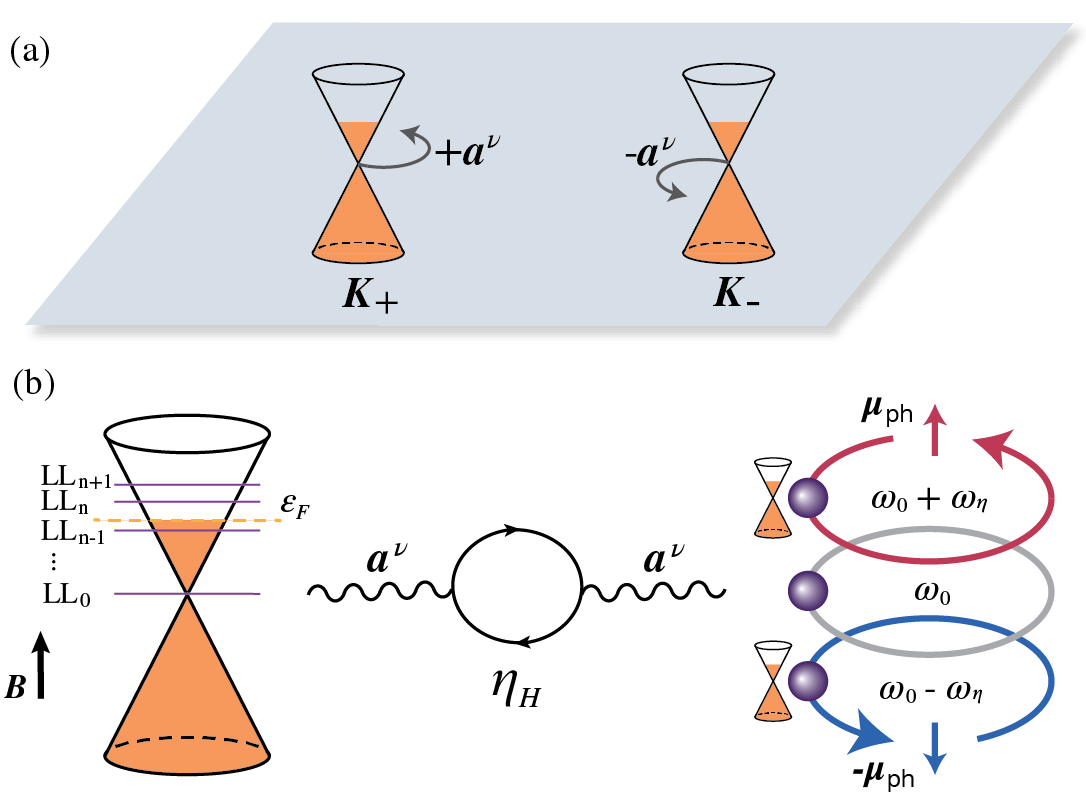}
    \caption{Schematics illustrating the mechanism for generating the giant orbital magnetic moment of phonons in doped Dirac semimetals. (a) Emergent gauge picture for the electron-phonon coupling in the Dirac semimetals. In momentum space, an emergent gauge field $\boldsymbol{a}^{\nu}$ induced by a phonon mode $\nu$ shifts the Dirac cones at two valleys oppositely. (b) Under magnetic fields, the Dirac fermions fill in Landau levels. The one-loop diagram for the phonon Hall viscosity $\eta_H$. Through $\eta_H$, the phonon magnetic moment is directly linked to the Hall conductivity and leads to frequency splitting in magnetic fields.}
    \label{fig:enter-label}
\end{figure}

\textit{Phonon Hall viscosity}.--- We move on to discuss the topological quantum field theory of the emergent gauge field. Integrating out the fermionic degrees of freedom in the system defined by Eq. (\ref{hamiltonian}), the quantum corrections from the Dirac fermions give rise to a Chern-Simons (CS) term in the effective action, if the gauge fields are (2+1)-dimensional \cite{Elastic_Gauge,gauge-theory}. For the EM gauge field $\boldsymbol{A}$, the CS term is $S_{\text{CS}}[\boldsymbol{A}]=\sigma_{xy}/2\int{d^3x \epsilon^{ijk} A_i \partial_j A_k}$. It describes a Hall conductivity response that reads $J_i \equiv -\delta S_{\text{CS}}[\boldsymbol{A}]/\delta A_j = \sigma_{xy}\epsilon_{ij}E_j$. The CS term can also be generalized to (3+1) dimesional Weyl semimetals, because the Weyl semimetals can be constructed by stacking Chern insulators in the momentum space \cite{PhysRevB.86.115133}. Since the emergent gauge field $\boldsymbol{a}^{\nu}$ couple to the Dirac fermions equivalently as $\boldsymbol{A}$, we write a CS term of the same form associated to $\boldsymbol{a}^{\nu}$,
\begin{equation}
    S_{\text{CS}}[\boldsymbol{a}^{\nu}] = \frac{\sigma_{xy}}{2}\int{d^3x} \epsilon^{ijk} a^{\nu}_i \partial_j a_k^{\nu}.
\end{equation}
The valley index $\chi$ is dropped out because $S_{\text{CS}}[\boldsymbol{a}^{\nu}]$ is quadratic in $\boldsymbol{a}^{\nu}$, which indicates the contributions from electrons at two valleys add up. We identify $S_{\text{CS}}[\boldsymbol{a}^{\nu}]$ as the effective term describing the phonon Hall viscosity \cite{Elastic_Gauge,Dissipationless-phonon}. This response is universal in topologically nontrivial systems since it is directly linked to the Hall conductivity $\sigma_{xy}$. The coefficient of phonon Hall viscosity $\eta_H$ is the antisymmetric part of a general viscosity tensor $\eta_{ijkl}$ \cite{Viscosity}. It is dissipationless and exists only when TRS is broken.

As we show next, the phonon Hall viscosity will modify the phonon dynamics. For the sake of simplicity, here we restrict ourselves to a model of phonons defined by the Lagrangian $\mathcal{L}_{\text{ph}} = (\rho_I/2)[(\dot{Q}^{\nu}_x)^2 + (\dot{Q}^{\nu}_y)^2 - (\omega_0^{\nu}Q^{\nu}_x)^2 - (\omega_0^{\nu}Q^{\nu}_y)^2]$, where $\rho_I$ is the ionic mass density, $Q^{\nu}_{x,y}$ are the linearly polarized phonon displacements, and $\omega_0^{\nu}$ is the $\Gamma$-point($\boldsymbol{q}=0$) frequency. This model describes doubly degenerate optical modes in the long-wavelength limit, but our theory applies generally to the modes that have the emergent gauge field description. We further assume $C_{4z}$ rotation symmetry and thus the emergent gauge field is simply $\boldsymbol{a}^{\nu} = (g^{\nu}/ev_F) \boldsymbol{Q}^{\nu}$. To switch on the phonon Hall viscosity $\eta_H$, an out-of-plane magnetic field $B$ is applied to break the TRS. As a result, there is an additional phonon viscosity term in the phonon Lagrangian,
\begin{equation}
    \mathcal{L}_{\eta} = \eta_H (Q^{\nu}_y \dot{Q}^{\nu}_x - Q^{\nu}_x \dot{Q}^{\nu}_y),
\end{equation}
where $\eta_H = \sigma_{xy}(g^{\nu})^2/(2e^2 v_F^2)$. We solve the equations of motion in the circularly polarized basis $\{Q_{l/r}^{\nu} = (Q_{x}^{\nu}\pm i Q_{y}^{\nu})/\sqrt{2}\}$. The frequencies of the left-handed and right-handed polarized modes are $\omega^{\nu}_{l/r} = [(\omega^{\nu}_0)^2 + (\eta_H/\rho_I)^2]^{1/2}\pm \eta_H/\rho_I$, respectively. We find a splitting of the phonon frequencies given by $\delta\omega^{\nu}  = 2\eta_H/\rho_I$ that is proportional to $\eta_H$. 

To obtain the magnetic field dependence of phonon frequencies, we now turn to compute the Hall conductivity. In Dirac semimetals, the Dirac-cone dispersion transforms into Landau levels (LLs) in a magnetic field. For a finite concentration $n_F$ of carriers (electrons or holes), the Fermi level $\varepsilon_F$ is between the $n$th and $(n+1)$th LL. Under relatively weak field, the LLs are filled to a large index ($|n|\gg$1) such that the Hall conductivity
response of the Dirac fermions is semiclassical. Hence we can use the Drude theory to calculate the Hall conductivity, $\sigma_{xy} = \sigma_{0}\omega_c\tau_{\text{tr}}/(1+\omega_c^2\tau_{\text{tr}}^2)$, where $\sigma_0$ is the dc conductivity, $\omega_c$ is the cyclotron frequency, and $\tau_{\text{tr}}$ is the transport lifetime. The field dependence enters through the cyclotron frequency given by $\omega_c = eB/m^*_c$ ($m^*_c$ is the cyclotron effective mass). As a result, we obtain the phonon frequencies as functions of $B$,
\begin{equation}
\label{freq_field_depend}
    \omega^{\nu}_{l/r} = \omega^{\nu}_0\left[\sqrt{1+\left(\xi\frac{\widetilde{\mu}_{\text{tr}} B}{1+\widetilde{\mu}_{\text{tr}}^2 B^2}\right)^2}\pm\xi\frac{\widetilde{\mu}_{\text{tr}} B}{1+\widetilde{\mu}_{\text{tr}}^2 B^2}\right]   
\end{equation}
where $\xi = {\sigma_0(g^{\nu})^2}/(2e^2 v_F^2\rho_I\omega_0^{\nu})$ and the carrier mobility $\widetilde{\mu}_{\text{tr}} = e\tau_{\text{tr}} / m_c^* $. In the quantum limit at large magnetic fields and low doping, one needs to go beyond the semiclassical approximation for $\sigma_{xy}$ and calculate $\sigma_{xy}$ quantum mechanically, i.e. using the Kubo formula \footnote{We consider the region where the phonon mode is not resonant with the excitations between the Landau levels in the quantum limit.}.

\textit{Giant phonon magnetic moment}.--- The most important finding in Eq. (\ref{freq_field_depend}) is that the frequencies of left-handed and right-handed polarized phonons split linearly with $B$ when the field is weak ($B\ll \widetilde{\mu}_{\text{tr}}^{-1}$). This linear field dependence is identified as the phonon Zeeman effect \cite{Dynamical-multiferroicity,Orbital-magnetic}. It is due to the magnetic moment of phonons interacting with the applied external magnetic field via the Zeeman coupling of the form $\hbar\omega^{\nu}_{l/r} = \hbar\omega^{\nu}_0\pm\boldsymbol{\mu}_{\text{ph}}\cdot\boldsymbol{B}$. We have
\begin{equation}
\label{phonon_moment}
     \mu_{\text{ph}} = \frac{(g^{\nu})^2}{v_F^2 \rho_I B} \frac{\hbar}{2e^2} \sigma_{xy}
\end{equation}
where $\sigma_{xy}(B\rightarrow0)=\sigma_0 \widetilde{\mu}_{\text{tr}}B$. This is our main result: the phonon magnetic moment $\mu_{\text{ph}}$ is directly linked to the electrical Hall conductivity $\sigma_{xy}$ through the phonon Hall viscosity $\eta_H$. Finally, we express $\mu_{\text{ph}}$ in the unit of Bohr magneton as $\mu_{\text{ph}} = \sigma_0 \tau_{\text{tr}} (e v_F)^{-2}(g^{\nu})^2\rho_I^{-1}(m_e/m_c^{*})\mu_B$ in the limit of $\omega_c \tau\ll 1$.

Our theory is also applicable to gapped systems with broken TRS. In (2+1) dimensions, this can be shown by adding a mass term $m_{\chi}v_F^2\sigma^z$ to  Eq.~\eqref{hamiltonian}. Following the same derivation, we find a splitting of phonon energies,
\begin{equation}
\label{gapped}
\hbar\delta\omega^{\nu} = \frac{(g^{\nu})^2}{v_F^2\rho_I}\frac{\hbar}{e^2}\sigma_{xy}.
\end{equation}
Here we set $\varepsilon_F$ inside the gap, and $\sigma_{xy}$ is the anomalous Hall conductance induced by the electronic Berry curvature of a filled band, with $\sigma_{xy} = (e^2/4\pi\hbar)(m_+ / |m_+| - m_- / |m_-|)$ \cite{Model-for-a-Quantum-Hall-Effect}.

Having established the general theory framework, we now apply it to concrete examples of Dirac semimetals.  In general, the $e$-ph coupling matrix element $g^{\nu}_{\alpha\beta}$ is computed based on the density functional perturbation theory (DFPT)~\cite{baroni2001phonons}.  However, since we are intersted in the $e$-ph coupling in the form of Eq.~\eqref{hamiltonian}, we adopt the frozen phonon approach.  We compare the Dirac cone shifted by the phonons with its equilibrium position in the $\bm{k}$-space. The displacement, as denoted by $\boldsymbol{K}_\chi - e \chi \boldsymbol{a}^{\nu}/\hbar$, provides a measure of the emergent gauge field.  Below, the electron band structure calculations are performed using the \textsc{Quantum ESPRESSO} package \cite{QUANTUM_ESPRESSO}. The phonon spectra and eigenvectors are calculated using DFPT for graphene, and the finite-displacement approach for Cd$_3$As$_2$, respectively. Details of numerical calculations can be found in Ref.~\cite{supplemental}.

\begin{figure}
    \centering
    \includegraphics[width=\linewidth]{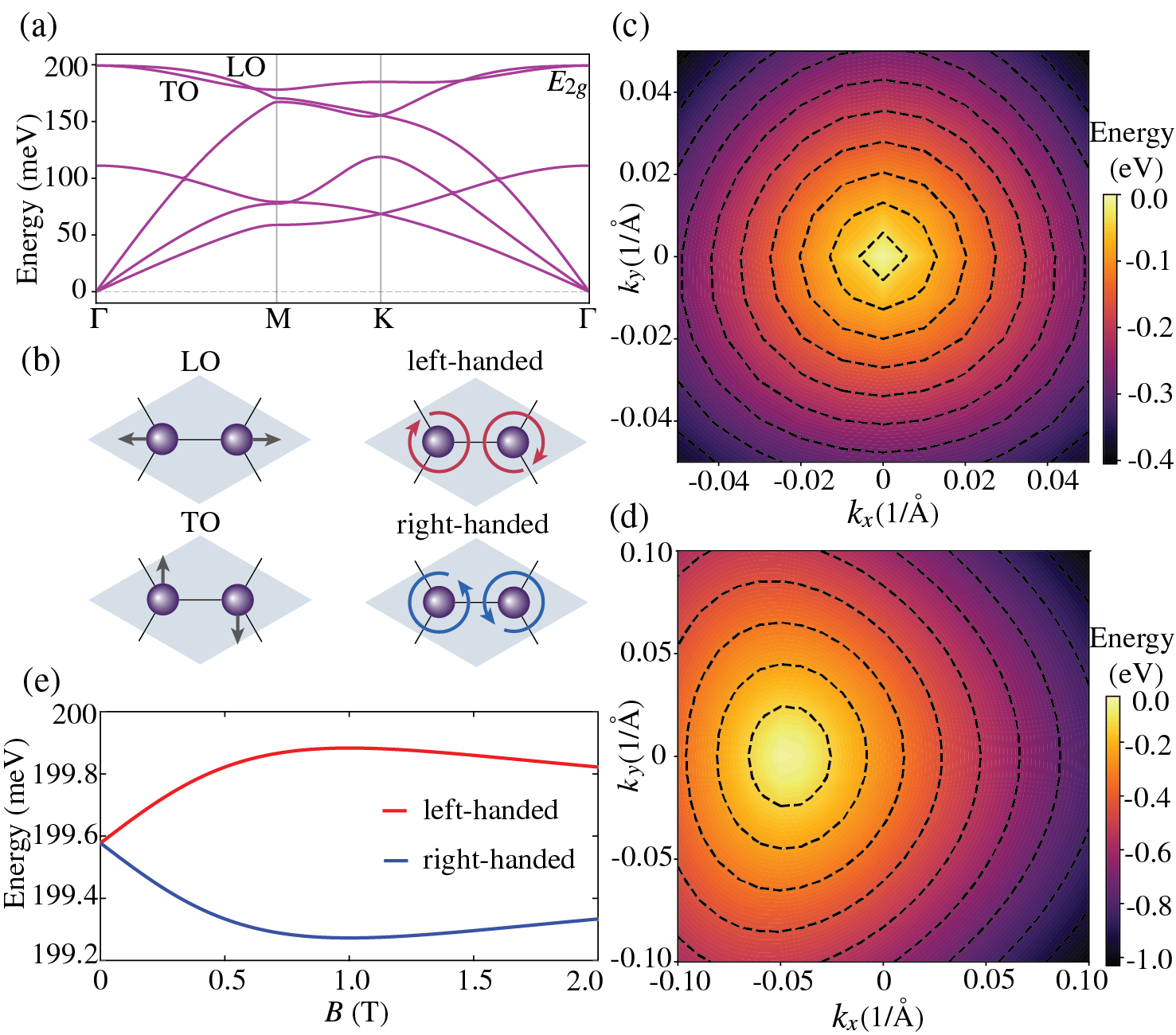}
    \caption{(a) Calculated phonon spectrum of monolayer graphene. (b) Vibrational modes at the $\Gamma$ point. The doubly-degenerate $E_{2g}$ pair consisting of LO and TO modes are used to construct the left-handed and right-handed circularly polarized phonons. (c)(d) Contours of the electronic band at the $\boldsymbol{K}_+$ valley in equilibrium and in the presence of the TO mode. (e) Phonon energy of the circularly polarized modes as functions of magnetic field. The linear splitting under weak magnetic fields is identified as the phonon Zeeman effect. }
    \label{fig:enter-label}
\end{figure}

We first focus on monolayer graphene, a material that hosts 2D massless Dirac fermions \cite{The-electronic}. Figure 2(a) shows the calculated phonon spectrum of graphene. At the $\Gamma$ point, there are two pairs of doubly-degenerate in-plane modes corresponding to the irreducible representations $E_{1u},E_{2g}$ of the $D_{6h}$ point group. We consider the Raman-active $E_{2g}$ pair, \textit{i.e.}, the $G$ band \cite{Raman-spectroscopy}, consisting of the longitudinal optical (LO) and transverse optical (TO) modes. As shown in Fig. 2(b), they can be used to construct left-handed and right-handed circular modes. The corresponding emergent gauge fields are ${a}^{E_{2g}}_{x} = (g^{\text{TO}}/ev_F){Q}^{\text{TO}}$ and ${a}^{E_{2g}}_{y} = -(g^{\text{LO}}/ev_F){Q}^{\text{LO}}$ \cite{Symmetry-based}. 

\begin{figure}
    \centering
    \includegraphics[width=\linewidth]{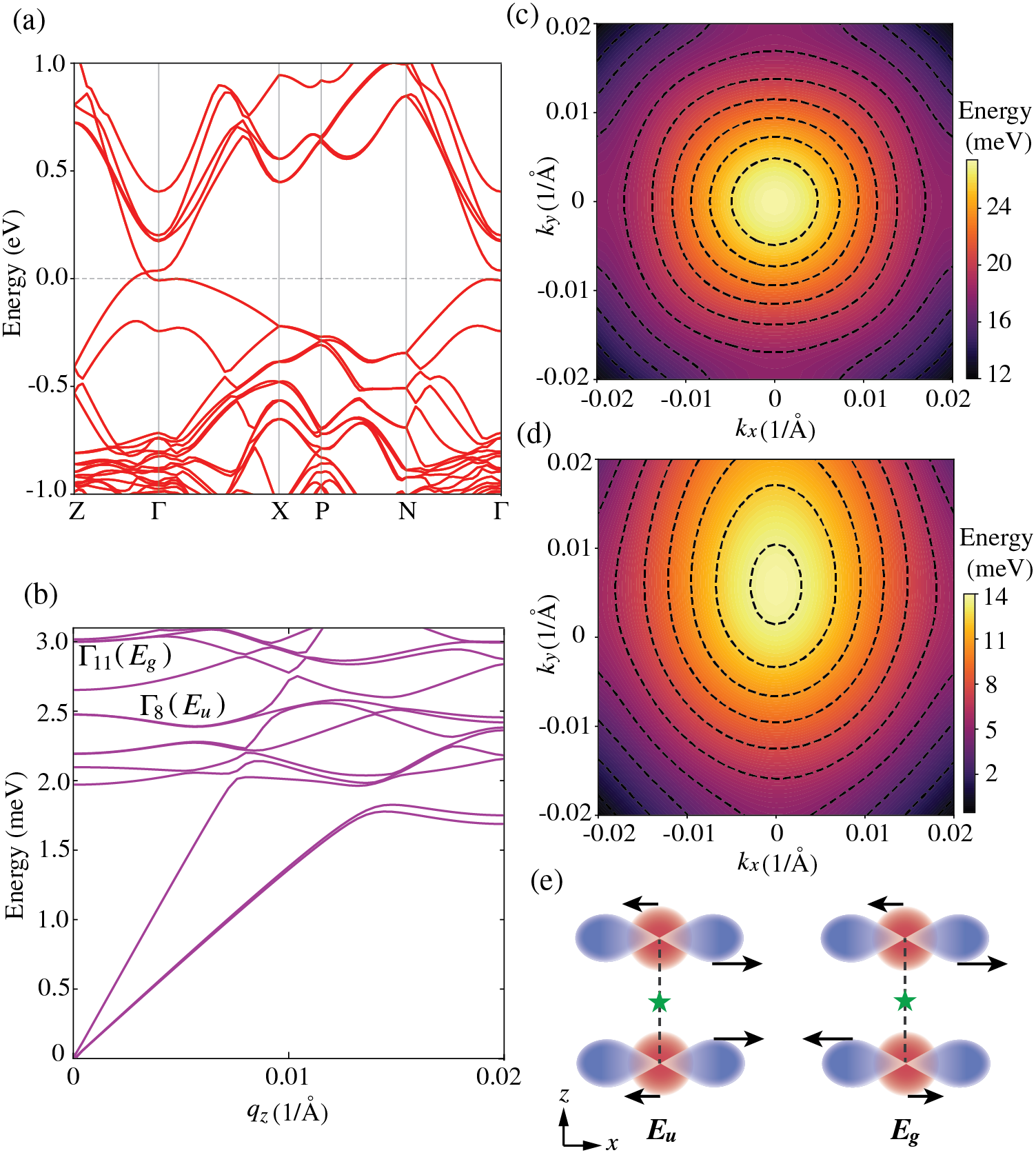}
    \caption{(a) Calculated electronic bands of $\text{Cd}_3\text{As}_2$ showing one Dirac point $\boldsymbol{K}_+$ on the $\Gamma$-Z axis. (b) Calculated phonon spectrum of $\text{Cd}_3\text{As}_2$ near the $\Gamma$-point along the $z$ direction. The $8$th and $11$th branches correspond to doubly-degenerate $E_u$ and $E_g$ optical modes, respectively. (c) (d) In-plane contours of the Dirac cone perpendicular to $\Gamma$-$Z$ at the $\boldsymbol{K}_+$ valley in equilibrium and in the presence of the $\Gamma_{11}$($E_g$) mode. The mode amplitude in (d) is $|\boldsymbol{Q}^{E_g}| = 0.75\ \text{\AA}$ and the shift distance of Dirac cone is $0.006\ \text{\AA}^{-1}$. (e) Illustrations of ion displacement associated with the optical phonons in the tight-binding model of $\text{Cd}_3\text{As}_2$. The star indicates an inversion center. Under the inversion operation, the $E_u$ ($E_g$) mode is antisymmetric (symmetric).}
    \label{fig:enter-label}
\end{figure}

Next we perform the first-principles simulation of $\boldsymbol{a}^{E_{2g}}$. Our calculations cover all three optical modes. While the out-of-plane mode does not couple to electrons due to the $\sigma_h$ symmetry, we find the LO and TO modes indeed induce gauge fields. In Fig. 2(c) and (d), electronic energy contours are plotted at the $\boldsymbol{K}_+$ valley in equilibrium and in the presence of the TO phonon mode. We see that the Dirac cone is shifted from its equilibrium position. From the shift distance $0.05 \ \text{\AA}^{-1}$ and the phonon displacement amplitude $0.03\ \text{\AA}$, the $e$-ph coupling matrix element $g^{\text{TO}}$ is $9.7\ \text{eV}\cdot \text{\AA}^{-1}$. To establish the validity of our method, we recalculate $g^{\text{TO}}$ directly using the DFPT \cite{supplemental}. The result $g^{\text{TO}}_{(\text{DFPT})} = 9.4 \ \text{eV}\cdot \text{\AA}^{-1}$ is in good agreement with our gauge field calculation, justifying the emergent gauge approach.

We adopt the following parameters for graphene \cite{Acoustically-Induced}: $v_F \approx 10^6\ \text{m}\cdot\text{s}^{-1}$, $\rho_I \approx 7.63\times10^{-7}\ \text{kg}\cdot\text{m}^{-2}$, $\widetilde{\mu}_{\text{tr}}\approx10^4\ \text{cm}^2\cdot\text{V}^{-1}\cdot \text{s}^{-1}$, $n_F\approx10^{11} \ \text{cm}^{-2}$. The field dependence of $\hbar\omega^{E_g}_{l/r}$ are plotted in Fig. 2(e) according to Eq. ($\ref{freq_field_depend}$). At $0.5$ Tesla, the splitting is $0.5\ \text{meV}$. The phonon magnetic moment given in Eq. (\ref{phonon_moment}) is calculated to be $\mu_{\text{ph}} = 10.6\ \mu_B$. Notably, a splitting of the $E_{2g}$ phonons in graphene under strong magnetic fields when the phonons are in resonance with the magnetoexcitons has been reported in the literature~\cite{Magneto-Raman-Scattering,Charge-Tuning-of-Nonresonant,Filling-Factor-Dependent,Electronic-properties-of-graphene}.  Our theory provides a non-resonance mechanism for phonon splitting based on the phonon Hall viscosity under weak magnetic fields. 

Finally, we turn to $\text{Cd}_3\text{As}_2$, an archetypal 3D Dirac semimetal. The electronic bands in Fig. 3(a) are calculated from the experimentally determined crystal structure with the space group $I4_1/acd$ \cite{The-Crystal}. Two 3D massless Dirac cones are located at momenta $\boldsymbol{K}_\chi = (0,0,\chi k_{z0})$. Figure 3(b) shows the phonon spectrum of $\text{Cd}_3\text{As}_2$ in the THz frequency range. We focus on the $\Gamma$-point in-plane optical phonons that are doubly degenerate. Depending on whether the modes are symmetric or antisymmetric under inversion, they belong to the irreducible representations $E_g, E_u$ of the point group $D_{4h}$, respectively. As showcase examples, we choose the $\Gamma_8$ branch with $E_u$ and the $\Gamma_{11}$ branch with $E_g$ in our first-principles calculations. We find the infrared-active $E_u$ mode does not induce a gauge field, forbidden by the inversion symmetry \cite{supplemental}. For the Raman-active $E_g$ mode, an emergent gauge field is allowed as plotted in the Fig. 3(d).

For an intuitive understanding, we develop a phonon-modulated tight-binding model to derive the emergent gauge field in $\text{Cd}_3 \text{As}_2$.
The low-energy electronic properties near the Fermi level can be described by an effective tetragonal lattice with $\text{Cd-}5s$ and $\text{As-}4p$ orbitals on each site \cite{Three-dimensional}. This model can capture the $e$-ph coupling of the acoustic modes \cite{Viscoelastic-response,Chiral-Anomaly}. To account for the optical modes, we formally double the unit cell and thus fold the Brillouin zone along $z$-axis. The ion displacement associated with the optical modes are shown in Fig. 3(e). Because of the relative rotation of $s$ and $p$ orbitals between neighboring sites, the inter-orbital hopping integrals along $z$ become nonzero. We find the modification to Hamiltonian from the $E_u$ mode does not contribute a gauge field, consistent with the first-principles calculations \cite{supplemental}.  For the $E_g$ mode, we obtain an effective $e$-ph coupling Hamiltonian in the basis of $\{c^\dagger_{s\uparrow,\boldsymbol{k}}, c^\dagger_{p\uparrow,\boldsymbol{k}},c^\dagger_{s\downarrow,\boldsymbol{k}}, c^\dagger_{p\downarrow,\boldsymbol{k}}\}\ket{0}$,
\begin{equation}
\begin{aligned}
    \delta \mathcal{H}_{\text{latt}}^{E_g} = \frac{t\beta}{l_z} \sin{k_zl_z} (Q^{E_g}_x\sigma^z\tau^x + Q^{E_g}_y\tau^y),
\end{aligned}
\end{equation}
where $t$ is the hopping integral, $\beta$ is the Grüneisen parameter, $l_z$ is the bond-length along $z$, and $\sigma_j,\tau_j$ are the Pauli matrices in spin and orbital space, respectively. Expanding at the valley $\boldsymbol{K}_\chi$ and comparing with the effective total Hamiltonian (\ref{hamiltonian}), we obtain the emergent gauge field: $ \boldsymbol{a}^{E_g} = (t\beta k_{z0}/e v_F)\boldsymbol{Q}^{E_g}$.

Due to the Zeeman interaction, the external magnetic field splits the Dirac points into Weyl nodes in $\text{Cd}_3 \text{As}_2$ \cite{Dirac-semimetal,Chiral-anomaly-factory}. The Weyl nodes are monopole sources of the Berry curvature that open a channel for intrinsic Hall conductivity. Hence our calculations of Hall conductivity need to include a Zeeman contribution given by $\sigma_{xy}^{\text{Zeeman}} = ({e^2}/{4\pi^2\hbar})\kappa_z$, where $\kappa_z$ is the distance between the Weyl nodes \cite{supplemental}. We adopt the following parameters for $\text{Cd}_3 \text{As}_2$ \cite{Ultrahigh-mobility-and-giant}: $v_F \approx 1.5\times10^6\ \text{m}\cdot \text{s}^{-1} $, $\rho_I \approx 3.03\ \text{g}\cdot\text{cm}^{-3}$, $n_F \approx 1.2 \times 10^{19}\ \text{cm}^{-3}$, $\widetilde{\mu}_{\text{tr}} \approx 3.2\times10^5\ \text{cm}^2\cdot\text{V}^{-1}\cdot\text{s}^{-1}$. The $e$-ph coupling matrix element $g^{E_g}$ from our first-principles calculations is $79.5\ \text{meV}\cdot\text{\AA}^{-1}$. The phonon magnetic moment is $\mu_{\text{ph}}=1.04\ \mu_B$. Interestingly, the order of magnitude of the calculated magnetic moment agrees well with the experiment \cite{A-Large-Effective}, despite that we considered Raman-active modes but the experiment measured the infrared-active mode. 

\textit{Conclusion and outlook.}--- We have established a theoretical framework to calculate the phonon magnetic moment in doped Dirac semimetals by treating phonon as an emergent gauge field. We find that the phonon magnetic moment is directly proportional to the Hall conductivity, indicating that a significant enhancement can be achieved with high carrier concentration and carrier mobility. Our theory is combined with the first-principles calculations, allowing us to quantitatively implement it to realistic materials. Magnetic moments are found to be on the order of Bohr magneton for the optical modes in graphene and $\text{Cd}_3 \text{As}_2$.
These modes are Raman active, and their magnetic moments can be measured by the phonon Zeeman splitting under magnetic fields using the Raman spectroscopy \cite{Observation-of-circularly,Magnetic-field-dependent-splitting,Truly-chiral}. Our results also pave the way for subsequent extensions to the infrared-active modes. In future experimental investigations, our theory offers tangible direction to search for large phonon magnetic moments in the topological quantum materials. 

\begin{acknowledgements}
{\it Acknowledgments.---} We thank Alexander Balatsky and Prashant Padmanabhan for the helpful discussion. The work at UW was supported by DOE Award No. DE-SC0012509. The work at LANL was carried out under the auspices of the U.S. DOE NNSA under contract No. 89233218CNA000001 through the LDRD Program, and was supported by the Center for Nonlinear Studies at LANL, and was performed, in part, at the Center for Integrated Nanotechnologies, an Office of Science User Facility operated for the U.S. DOE Office of Science, under user proposals $\#2018BU0010$ and $\#2018BU0083$.
\end{acknowledgements}

%

\pagebreak
\widetext
\begin{center}
\textbf{\large Supplemental materials for "Gauge theory of giant phonon magnetic moment in doped Dirac semimetals"}
\end{center}
\setcounter{equation}{0}
\setcounter{figure}{0}
\setcounter{table}{0}
\setcounter{page}{1}
\makeatletter
\renewcommand{\theequation}{S\arabic{equation}}
\renewcommand{\thefigure}{S\arabic{figure}}
\renewcommand{\bibnumfmt}[1]{[S#1]}
\renewcommand{\citenumfont}[1]{S#1}

\section{Electron-phonon coupling}
We start from the many-particle Hamiltonian of electrons and ions at equilibrium positions, $H = H_{\text{e}} + H_{\text{ion}} + H_{\text{ei}}$, where $H_{\text{ei}} = \sum_{i\boldsymbol{R}\alpha} V_{\text{ei}} (\boldsymbol{r}_i-\boldsymbol{R}-\boldsymbol{\tau}_\alpha)$ is the electron-ion interaction Hamiltonian. Here $\boldsymbol{R}$ is the lattice vector for unit cells and $\boldsymbol{\tau}_\alpha$ is the position of ion $\alpha$ within a unit cell. If the ions vibrate, $H_{\text{ei}}$ will change to $\sum_{i\boldsymbol{R}\alpha} V_{\text{ei}} (\boldsymbol{r}_i-\boldsymbol{R}-\boldsymbol{\tau}_\alpha + \delta\boldsymbol{\tau}_{\boldsymbol{R}\alpha})$, where $\delta\boldsymbol{\tau}_{\boldsymbol{R}\alpha}$ is the atomic displacement of the ion $\alpha$ in the unit cell $\boldsymbol{R}$ from its equilibrium position. It is usually small compared to the lattice constant, allowing us to expand the potential to the first order of $\delta\boldsymbol{\tau}_{\boldsymbol{R}\alpha}$ as $\sum_{i\boldsymbol{R}\alpha} V_{\text{ei}}(\boldsymbol{r}_i - \boldsymbol{R} - \boldsymbol{\tau}_\alpha) + \delta\boldsymbol{\tau}_{\boldsymbol{R}\alpha} \cdot \boldsymbol{\nabla} V_{\text{ei}}(\boldsymbol{r}_i - \boldsymbol{R} - \boldsymbol{\tau}_\alpha)+ O(\delta\boldsymbol{\tau}_{\boldsymbol{R}\alpha}^2)$. Thus the electron-phonon coupling Hamiltonian comes from the first-order correction,
\begin{equation}
    H_{e\text{-ph}} = \sum_{i\boldsymbol{R}\alpha} \delta\boldsymbol{\tau}_{\boldsymbol{R}\alpha} \cdot \boldsymbol{\nabla} V_{\text{ei}}(\boldsymbol{r}_i - \boldsymbol{R} - \boldsymbol{\tau}_\alpha) = \int d\boldsymbol{r} \rho(\boldsymbol{r}) \sum_{\boldsymbol{R}\alpha} \delta\boldsymbol{\tau}_{\boldsymbol{R}\alpha} \cdot \boldsymbol{\nabla} V_{\text{ei}}(\boldsymbol{r}_i - \boldsymbol{R} - \boldsymbol{\tau}_\alpha),
\end{equation}
where $\rho(\boldsymbol{r})$ is the density operator of electrons, and $\rho(\boldsymbol{r}) = \sum_{i} \delta(\hat{\boldsymbol{r}}_i - \boldsymbol{r})$ in the coordinate representation. Now we move to the second quantization. The recipe is using the second quantization form of density operator $\rho(\boldsymbol{r}) = \psi^\dagger(\boldsymbol{r})\psi(\boldsymbol{r})$. The field operator projected to a specific basis, such as Bloch electronic basis, is $\psi^\dagger(\boldsymbol{r}) =  \sum_{n\boldsymbol{k}}\psi^*_{n\boldsymbol{k}}(\boldsymbol{r}) c^\dagger_{n\boldsymbol{k}}$, where $n$ is the energy band index and $\boldsymbol{k}$ is electronic wavevector. The Bloch wavefunctions are given by $\psi_{n\boldsymbol{k}}(r) = u_{n\boldsymbol{k}}(\boldsymbol{r})e^{i\boldsymbol{k}\cdot\boldsymbol{r}}$, where $u_{n\boldsymbol{k}} = \braket{\boldsymbol{r}}{n\boldsymbol{k}}$ is the periodic Bloch function within one unit cell. Under this representation, the electron-phonon coupling Hamiltonian becomes
\begin{equation}
    H_{e\text{-ph}} = \sum_{nm} \sum_{\boldsymbol{k}\boldsymbol{k}'} \int d\boldsymbol{r} \psi^*_{n\boldsymbol{k}}(\boldsymbol{r}) \sum_{\boldsymbol{R}\alpha} \delta\boldsymbol{\tau}_{\boldsymbol{R}\alpha} \cdot \boldsymbol{\nabla} V_{\text{ei}}(\boldsymbol{r}_i - \boldsymbol{R} - \boldsymbol{\tau}_\alpha) \psi_{m\boldsymbol{k}'}(\boldsymbol{r})  c^\dagger_{n\boldsymbol{k}}c_{m\boldsymbol{k}'}.
\end{equation}
Now we expand the atomic displacement of the ion at $\boldsymbol{R}+\boldsymbol{\tau}_\alpha$ in terms of the vibrational modes,
\begin{equation}
    \label{disp_expsn}\delta\boldsymbol{\tau}_{\boldsymbol{R}\alpha} = \sum_{\boldsymbol{q},\nu} \sqrt{\frac{\overline{M}}{NM_{\alpha}}}e^{i\boldsymbol{q} \cdot \boldsymbol{R}}\ \boldsymbol{\xi}_{\alpha}^{(\nu)}(\boldsymbol{q}) Q_{\boldsymbol{q}}^{(\nu)},
\end{equation}
where $\boldsymbol{q}$ is the phonon wavevector, $\nu$ labels the normal modes, $\omega_{\boldsymbol{q}}^{(\nu)}$ is the frequency dispersion, $\boldsymbol{\xi}_{\alpha}^{(\nu)}(\boldsymbol{q})$ is the normal mode eigenvector or polarization vector that solves the dynamic matrix problem, $\overline{M} = 1/N_\alpha\sum_\alpha M_\alpha$ is the average mass of $N_\alpha$ atoms in a unit cell, $Q_{\boldsymbol{q}}^{(\nu)}$ refers to as the complex normal coordinate of the displacement projected onto the normal mode $\nu$. Since the left hand side of equation (\ref{disp_expsn}) is real and the normal mode eigenvectors satisfy $\boldsymbol{\xi}_{\alpha}^{(\nu)}(\boldsymbol{q})^* = \boldsymbol{\xi}_{\alpha}^{(\nu)}(-\boldsymbol{q}) $, we obtain the complex conjugation relation, $Q_{\boldsymbol{q}}^{(\nu)*} = Q_{-\boldsymbol{q}}^{(\nu)}$. The quantization of the phonons is given by 
\begin{equation}
    Q_{\boldsymbol{q}}^{(\nu)} = \sqrt{\frac{\hbar}{2\overline{M}\omega^{(\nu)}_{\boldsymbol{q}}}} [a^{(\nu)}_{\boldsymbol{q}} + a^{(\nu)\dagger}_{-\boldsymbol{q}}],
\end{equation}
where $M_{\alpha}$ is the mass of the ion $\alpha$, $\omega_{\boldsymbol{q}}^{(\nu)}$ is the frequency of mode $\nu$, and $a^\dagger(a)$ is the creation (annihilation) operator of phonons. We arrives at the Frölich Hamiltonian for the electron-phonon coupling,
\begin{equation}
\begin{aligned}
  H_{e\text{-ph}} = \frac{1}{\sqrt{N}}\sum_{nm\boldsymbol{k}} \sum_{\boldsymbol{q},\nu} g_{mn}^{(\nu)}(\boldsymbol{k},\boldsymbol{q}) Q_{\boldsymbol{q}}^{(\nu)}  c^\dagger_{n\boldsymbol{k}+\frac{\boldsymbol{q}}{2}}c_{m\boldsymbol{k}-\frac{\boldsymbol{q}}{2}},
\end{aligned}
\end{equation}
where the electron-phonon coupling matrix element is given by
\begin{equation}
    g_{mn}^{(\nu)}(\boldsymbol{k},\boldsymbol{q}) =  \bra{n\boldsymbol{k}+\frac{\boldsymbol{q}}{2}} ({\nabla}V_{\text{ei}})_{\boldsymbol{q}}^{(\nu)} \ket{m\boldsymbol{k}-\frac{\boldsymbol{q}}{2}}_{\text{uc}}.
\end{equation}
We have defined $({\nabla}V_{\text{ei}})_{\boldsymbol{q}}^{(\nu)} = \sum_{\boldsymbol{R}\alpha}\sqrt{\overline{M}/M_{\alpha}} e^{-i\boldsymbol{q} \cdot (\boldsymbol{r} - \boldsymbol{R})} \boldsymbol{\xi}_{\alpha}^{(\nu)}(\boldsymbol{q}) \cdot \boldsymbol{\nabla}V_{\text{ei}}(\boldsymbol{r} - \boldsymbol{R}-\boldsymbol{\tau}_\alpha)$. Bloch thereom has been used. Evaluating $g_{mn}^{(\nu)}(\boldsymbol{k},\boldsymbol{q})$ is the essential part in the calculations of the electron-phonon coupling. Common approaches are the density functional perturbation theory and the frozen phonon approach, which we will employ. Other approaches such as the tight-binding method are useful for different purposes. 

\section{Phonon-modulated tight-binding model for $\text{Cd}_3\text{As}_2$}
In this section, we develop a phonon-modulated tight-binding model for $\text{Cd}_3\text{As}_2$. We start from the model that captures the band inversion of the atomic Cd-$5s$ and As-$4p$ near the $\Gamma$ point \cite{Three-dimensional}. The basis states are $\ket{S_{\frac{1}{2}}\frac{1}{2}},\ket{P_{\frac{3}{2}}\frac{3}{2}},\ket{S_{\frac{1}{2}}-\frac{1}{2}},\ket{P_{\frac{3}{2}}-\frac{3}{2}}$. The Hamiltonian is 
\begin{equation}
\label{k.p ham}
    H(\boldsymbol{k}) = \varepsilon_0(\boldsymbol{k}) \hat{I} + M(\boldsymbol{k})\tau^z + A k_x \sigma^z\tau^x + A k_y \tau^y,
\end{equation}
where $\varepsilon_0(\boldsymbol{k}) = C_0 + C_1 k_z^2 + C_2(k_x^2 + k_y^2)$, $M(\boldsymbol{k}) = M_0 + M_1 k_z^2 + M_2(k_x^2 + k_y^2)$ with parameters $C_{0,1,2}$, $M_{0,1,2}$, and $A$ fitted from the DFT calculations, and $\boldsymbol{\sigma},\boldsymbol{\tau}$ are Pauli matrices in spin and orbital space. Since spin-up and spin-down blocks are effectively decoupled in the model Hamiltonian, we can analyze them separately. In the following, we focus on the spin-up block that is based on $\ket{S_{\frac{1}{2}}\frac{1}{2}},\ket{P_{\frac{3}{2}}\frac{3}{2}}$,
\begin{equation}
    h = M(\boldsymbol{k})\tau^z + A k_x\tau^x + A k_y \tau^y.
\end{equation}
By making the substitution,
\begin{equation}
    \begin{aligned}
        k_{x,y}&\rightarrow\frac{1}{a}\sin{(k_{x,y}a)},\\
        k_{x,y,z}^2&\rightarrow\frac{2}{a^2}[1-\cos{(k_{x,y,z}a)}],
    \end{aligned}
\end{equation}
we regularize the Hamiltonian on a simple cubic lattice with lattice constant $a$,
\begin{equation}
\label{latt_ham}
    h^{\text{latt}} = [m_0 + m_1\cos{(k_za)}+m_2\cos{(k_xa)}+m_2\cos{(k_ya)}]\tau^z + t\sin{(k_xa)}\tau^x + t\sin{(k_ya)}\tau^y,
\end{equation}
where $m_0 = M_0 + \frac{2M_1}{a^2} + \frac{4M_2}{a^2}$, $m_1 = -\frac{2M_1}{a^2}$, $m_2 = -\frac{2M_2}{a^2}$, $t=\frac{A}{a}$. The Hamiltonian (\ref{latt_ham}) has valleys at $\boldsymbol{K}_\chi = (0,0,\chi k_{z0})$, where $\chi = \pm 1$ and $k_{z0}$ is given by $\cos(k_{z0}a) = -(m_0 + 2m_2) / m_1$. At the valleys, we expand $h^{\text{latt}}(\boldsymbol{K}_{\chi}+\boldsymbol{k})$ in $\boldsymbol{k}$ to obtain $h_{\chi} = \hbar v^j_{\chi} k_j \tau^j$ with the velocity vector $\boldsymbol{v}_{\chi} = \hbar^{-1}a(t,t,-\chi m_1 \sin{(k_{z0}a)})$. Although real $\text{Cd}_3\text{As}_2$
crystal has a complex structure with 80 atoms per unit cell, its low-energy physics can be well described by this effective tight-binding model with Cd-$5s$ and As-$4p$ orbitals on the vertices of the cubic lattice. The real-space tight-binding Hamiltonian is given by $H=\sum_j (h_j^{\text{intra}} + h_j^{\text{inter}})$ where
\begin{equation}
    \begin{aligned}
        h_j^{\text{intra}} &= m_0(a^\dagger_j a_j - b^\dagger_j b_j)\\
        &+\frac{m_1}{2}(a^\dagger_j a_{j+a\hat{x}} - b^\dagger_j b_{j+a\hat{x}} + a^\dagger_j a_{j+a\hat{y}} - b^\dagger_j b_{j+a\hat{y}}) + h.c.\\
        &+\frac{m_2}{2}(a^\dagger_j a_{j+a\hat{z}} - b^\dagger_j b_{j+a\hat{z}}) + h.c.\\
    \end{aligned}
\end{equation}
and
\begin{equation}
    \begin{aligned}
        h_j^{\text{inter}} &= -i\frac{t}{2}(a^\dagger_j b_{j+a\hat{x}} + b^\dagger_j a_{j+a\hat{x}}) + h.c.\\
        &-\frac{t}{2}(a^\dagger_j b_{j+a\hat{y}} + b^\dagger_j a_{j+a\hat{y}}) + h.c..
    \end{aligned}
\end{equation}
Here, $a_j(a^\dagger_j)$ annihilates (creates) a fermion in the Cd-$5s$ orbital at site $j$ and $b_j(b^\dagger_j)$ annihilates (creates) a fermion in the As-$4p$ orbital at site $j$.
In order to capture the in-plane doubly degenerate $E_u$ and $E_g$ optical modes, we need at least two Cd atoms and two As atoms in a unit cell. We formally double the lattice unit cell along the $z$ direction as shown in Fig. 3(e). 

The effect of phonons on the tight-binding Hamiltonian is to modify the hopping integrals. The correction has two types: the bond-length change, and the rotations between orbitals. The bond-length change is isotropic and exists for all orbitals \cite{Elastic_Gauge,Viscoelastic-response,Chiral-Anomaly}. If more than one orbital (or local degree of freedom) is present in each unit cell, there can be a correction to the hopping term between unlike orbitals due to a relative rotation seen from neighboring sites. 

For our purpose of deriving the emergent gauge field that shifts the Dirac cones on the $\Gamma$-$Z$ axis, the most important modification is the change of the hopping integrals along $z$. Because of the relative rotation of $s$ and $p$ orbitals between neighboring sites, the inter-orbital hopping integrals become nonzero along $z$ \cite{Viscoelastic-response},
\begin{equation}
    t_{sp}(a \hat{z} + \delta\boldsymbol{\tau}) \approx \frac{\boldsymbol{n}\cdot(a\hat{z}\times\delta\boldsymbol{\tau})}{a} \frac{\partial t_{sp} (a\hat{x})}{\partial a},
\end{equation}
where $\boldsymbol{n}$ is the normal vector to the plane and $t_{sp}(a\hat{x})$ is the hopping amplitude between $s$ and $p$ orbitals in the $x$ direction. The displacement $\delta\boldsymbol{\tau}$ can be related the phonon displacement using Eq. (\ref{disp_expsn}). For the $E_u$ mode $\Gamma_{8}$, the modification to the Hamiltonian at site $j$ is 
\begin{equation}
    \delta h_j^{\Gamma_8} = -i \frac{t}{2}\frac{\beta Q_x^{E_u}}{a}[a^\dagger_{j} b_{j+a\hat{z}} + a^\dagger_{j} b_{j-a\hat{z}} + b^\dagger_{j} a_{j+a\hat{z}} + b^\dagger_{j}a_{j-a\hat{z}}] + h.c.
\end{equation}
where $\beta = \frac{a}{t}\frac{\partial t}{\partial a}$ is the Grüneisen parameter of the model. We have $\sum_j \delta h_j^{\Gamma_8}=0$. For the $E_g$ mode $\Gamma_{11}$, 
\begin{equation}
    \delta h_j^{\Gamma_{11}} = -i \frac{t}{2}\frac{\beta Q_x^{E_g}}{a}[a^\dagger_{j} b_{j+a\hat{z}} - a^\dagger_{j} b_{j-a\hat{z}} + b^\dagger_{j}a_{j+a\hat{z}} - b^\dagger_{j}a_{j-a\hat{z}}] + h.c.
\end{equation}
and the $k$-space Hamiltonian is $\delta h^{\Gamma_{11}}(\boldsymbol{k}) = t\frac{\beta Q_x^{E_g}}{a}\sin{(k_za)}\sigma^x$. Similar analysis can be applied to the $y$-polarized mode. Therefore, we obtain an effective Hamiltonian that captures the correction from the $e$-ph coupling with the $E_g$ mode,
\begin{equation}
    \delta {H}_{\text{latt}}^{E_g}(\boldsymbol{k}) = \frac{t\beta}{a} \sin{(k_za)} (Q^{E_g}_x\sigma^z\tau^x + Q^{E_g}_y\tau^y).
\end{equation}
Expanding in the vicinity of $K_{\chi}$, we obtain the phonon-modulated effective total Hamiloninian,
\begin{equation}
\label{eff_ham}
    {H}_{\text{eff}}(\boldsymbol{k}) =  v_F  (\hbar k_x - \frac{t \beta  k_{z0}}{v_F} \chi Q^{E_g}_x) \sigma^z \tau^x + v_F  (\hbar k_y - \frac{t \beta  k_{z0}}{v_F} \chi Q^{E_g}_y) \tau^y + \chi v_z \hbar k_z\tau^z,
\end{equation}
where $v_F = \hbar^{-1} a t$ and $v_z = \hbar^{-1} a m_1 \sin{(k_{z0}a)}$. 
These results are consistent with the symmetry analysis: the emergent gauge field description must be compartible with the little group at the Dirac nodes. The little group at the Dirac node $\boldsymbol{K}_\chi$ for the model Eq. \eqref{k.p ham} contains $C_{4z}$ and $T\times I$, where $C_{4z}$ is the 4-fold ration along the $z$ axis, $T$ is the time reversal and $I$ is the inversion transformation. It is obvious that both $\mathbf{k}$ and $\mathbf{Q}^{\nu}$ transform in the same way under $C_{4z}$.
Under $T\times I$, the electronic momentum $\hbar \boldsymbol{k}$ is even. In order to allow for gauge description, the $\chi \boldsymbol{Q}^{\nu}$ term must transform in the same way as $\hbar \boldsymbol{k}$ that is even. Since the valley index $\chi$ is even under $T\times I$, $\boldsymbol{Q}^{\nu}$ must also be even. As a result, the inversion-odd $E_u$ mode is forbidden to couple as an emergent gauge field, while the inversion-even $E_g$ mode is allowed as shown in the Eq. (\ref{eff_ham}).  

\section{Hall conductivity of $\text{Cd}_3\text{As}_2$}

We consider an external magnetic field applied along the $z$-axis of $\text{Cd}_3\text{As}_2$. The field couples into the Hamiltonian in two distinct ways, ${H}\to{H}(\boldsymbol{p}+e\boldsymbol{A}) + {H}_Z$. The first part is the orbital effect, and the second part is the Zeeman effect. Thus, there are
two contributions to Hall conductivity. The orbital contribution is calculated in the semiclassical limit using the Drude model,
\begin{equation}
  \sigma_{xy}^{\text{orbital}} = \sigma_0\frac{\omega_c\tau}{1 + \omega_c^2\tau^2}.
\end{equation}
The Zeeman field breaks the time-reversal symmetry and splits the Dirac points into Weyl points in momentum space and also causes a shift in energy \cite{Chiral-anomaly-factory, Dirac-semimetal}. We consider the Hamiltonian in the Eq. (\ref{k.p ham}). The Zeeman term takes the form
\begin{equation}
    H_{Z} = - (g_s \boldsymbol{J}_s + g_p \boldsymbol{J}_p) \cdot \boldsymbol{B},
\end{equation}
where $g_s$ and $g_p$ are $g$-factors for the $s$ and $p$ orbitals. When $\boldsymbol{B} = B_z\hat{z}$, the total Hamiltonian is 
\begin{equation}
\begin{aligned}
    H = \epsilon_0(k) + \begin{bmatrix}
        M(k) - \frac{3}{2} g_p B_z & A(k) & 0 & 0 \\
        A^*(k) & -M(k)- \frac{1}{2} g_s B_z & 0 & 0 \\
        0 & 0 & -M(k) + \frac{1}{2} g_p B_z & -A(k)\\
        0 & 0 & -A^*(k) & M(k) + \frac{3}{2} g_p B_z
    \end{bmatrix}.
\end{aligned}
\end{equation}
The details of solving this Hamiltonian can be found in Ref. \cite{Chiral-anomaly-factory}. For any magnitude of $B_z$, at least one pair of Weyl points exist; for small enough $B_z\ll|\frac{4 M_0}{g_s - 3g_p}|$, both pairs exist. By adopting the parameters $M_0 = 0.02\  \text{eV}$, $g_s = 18.6$, $g_p = 2$ for $\text{Cd}_3\text{As}_2$, the critical value of $B_z$ is calculated as $\approx 50\ T$. Thus, we consider the case where four Weyl nodes exist and given by
\begin{equation}
    k_{z0} = \pm \sqrt{\frac{\pm(g_s - 3g_p)B_z/4-M_0}{M_1}}.
\end{equation}
The Weyl nodes are monopole sources of the Berry curvature that open a channel for intrinsic Hall conductivity. Hence, our calculations of Hall conductivity need to include a Zeeman contribution given by
\begin{equation}
    \sigma_{xy}^{\text{Zeeman}} = \frac{e^2}{4\pi^2\hbar}\kappa_z,
\end{equation}
where $\kappa_z$ is the generalized distance between the four Weyl nodes, which is the $z$-component of the vector defined by
\begin{equation}
    \boldsymbol{\kappa} = \sum_i C_i \boldsymbol{P}_i = \sum_i (-1)^{\xi^i} \boldsymbol{P}_i.
\end{equation}
Here, $i$ labels four different nodes, $C_i$ are their topological charges, $\boldsymbol{P}_i$ are their momenta, and $\xi^i$ are their chiralities. When the magnetic field is weak, the Zeeman-induced shift of momenta is small compared to the momenta of the Weyl nodes, allowing us to expand $\kappa_z$ to the linear order in $B_z$ as $\kappa_z = \sqrt{-M_0/M_1} (g_s - 3 g_p)\mu_B B_z / M_0$, where we have restored the Bohr magneton $\mu_B$. For the parameters $M_0 = 0.02\  \text{eV}$, $M_1 = -18.77\  \text{eV}\cdot\text{\AA}^2$, $g_s = 18.6$, $g_p = 2$, and $B_z = 1\ T$,  the Zeeman contribution to Hall conductivity is $\sigma_{xy}^{\text{Zeeman}}\approx (7.5\times 10^{-3}{\text{\AA}}^{-1})\frac{e^2}{h}$. For comparison, the orbital contribution at $B_z = 1\ T$ calculated using the parameters in the main text is $\sigma_{xy}^{\text{orbital}}\approx (2\ \text{\AA}^{-1})\frac{e^2}{\hbar}$, which dominates over the Zeeman contribution.

\section{First-principles calculations}
In this section, we present numerical details on the electron band structures, phonon spectra, and electron-phonon coupling. 

All DFT calculations are performed within the \textsc{Quantum Espresso} package~\cite{QUANTUM_ESPRESSO}. For graphene, 
we utilize the optimized norm-conserving Vanderbilt (ONCV) pseudopotential~\cite{hamann2013optimized} and the local density functional. An energy cut-off of 90 Ry is used to expand the wave functions. Phonon dispersions and eigenvectors are calculated through the DFPT using the \textsc{Quantum Espresso} package. Since the TO and LO modes are degenerate at the $\Gamma$ point, we can construct two linearly polarized modes that are along the $x$ and $y$ directions, respectively.
The atoms are then displaced along the two modes with a phonon amplitude of 0.03 \AA.

\begin{figure}
\setcounter{figure}{0}
\renewcommand{\figurename}{Fig.}
\renewcommand{\thefigure}{S\arabic{figure}}
    \centering
    \includegraphics[width=0.8\linewidth]{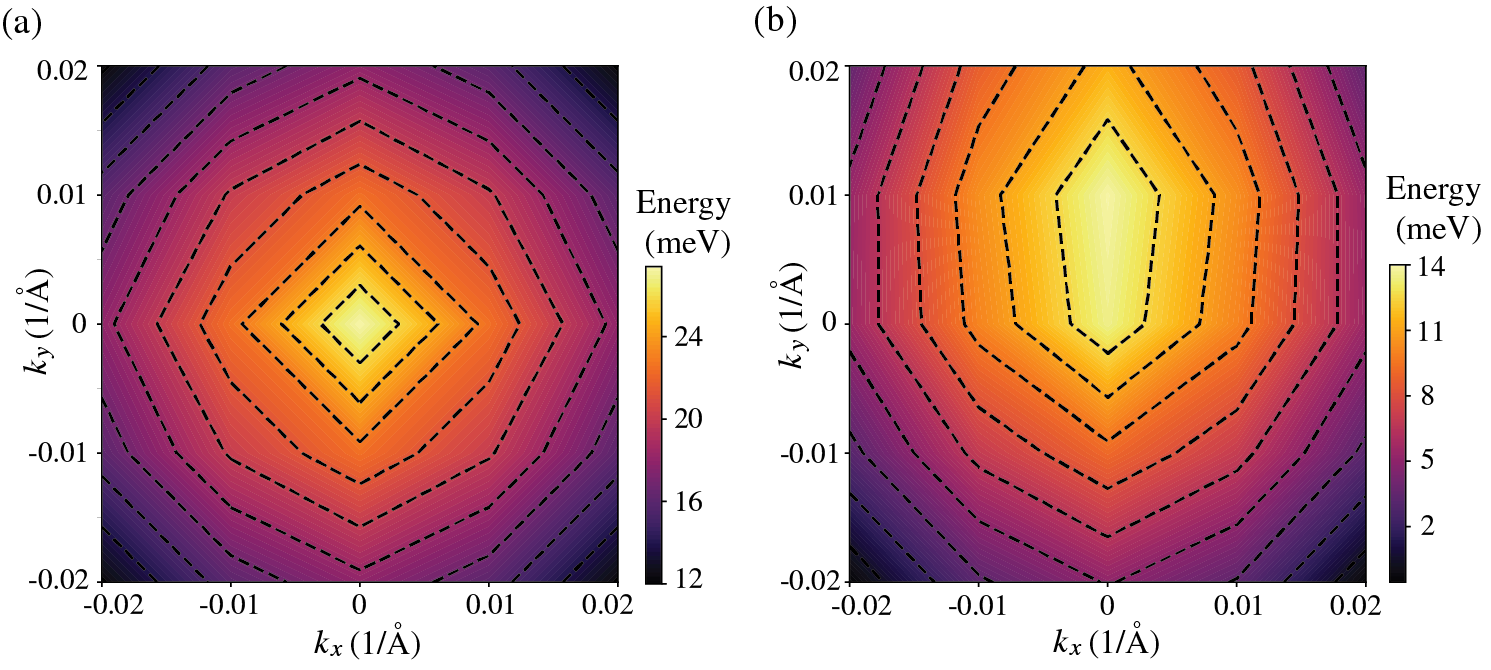}
    \caption{In-plane contour plots of the Dirac cone perpendicular to the $\Gamma$-$Z$ direction on a $k$-grid of $5\times5$ calculations using the DFT in equilibrium and in the presence of the $\Gamma_{11}$($E_g$) mode. }
    \label{fig:enter-label}
\end{figure}

For Cd$_3$As$_2$, we employ the optimized norm-conserving Vanderbilt (ONCV) pseudopotential~\cite{hamann2013optimized} and the Perdew-Burke-Ernzerhof (PBE) functional~\cite{perdew1996generalized}. 
Atomic structural parameters are obtained from Ref.~\cite{The-Crystal}.
An energy cut-off of 100 Ry and a $k$-grid of $2\times2\times2$ are used for self-consistent calculations. Phonon dispersions and eigenvectors are calculated based on the finite-displacement approach using the \textsc{Phononpy} package~\cite{togo2023first,togo2023implementation}. Similar to graphene, we construct two linearly polarized modes from the doubly degenerate E$_g$ modes at the $\Gamma$ point and displace atoms along the phonon eigenvectors with an amplitude of 0.75 \AA. Due to the considerable computational cost, we initially use a $k$-grid of $5\times5$ in DFT calculations to obtain the in-plane contour plot of the Dirac cone perpendicular to the $\Gamma$-$Z$ direction as in Fig. S1 (a) and (b), subsequently interpolating the map to a $41\times41$ $k$-grid.

\end{document}